\documentclass{aa}  

\usepackage{graphicx}
\usepackage{txfonts}
\usepackage{lipsum}
\usepackage{subcaption}         
\usepackage{lscape}             
\usepackage{placeins}           
                                
\usepackage{xcolor}
\usepackage{hyperref}
\hypersetup{
    colorlinks=true,
    linkcolor=blue,    
    filecolor=blue,    
    urlcolor=blue,     
    citecolor=blue     
}

\def \nustar {\mbox{\emph{NuSTAR}}}
\def \xmm {\mbox{\emph{XMM-Newton}}}

\newcommand{\swift}{\textit{Swift}}
\newcommand{\integral}{INTEGRAL}

\newcommand {\beq}{\begin {eqnarray}}
\newcommand {\eeq}{\end {eqnarray}}

\newcommand{\src}{IGR\,J00370+6122}

\begin{document}

%
\title{A bright X-ray flaring event from the intermediate supergiant fast X-ray transient IGR J00370+6122 observed with NuSTAR}
   \titlerunning{X-ray flares from the intermediate SFXT IGR J00370+6122}
   
%
%
%

      \author{L.~Ducci\inst{\ref{tue}}\fnmsep\corrauth{ducci@astro.uni-tuebingen.de}
        \and M.~M.~Serim\inst{\ref{metu}}\email{mserim@metu.edu.tr}
        \and P.~Romano\inst{\ref{oab}}\email{patrizia.romano@inaf.it}
        \and E.~Bozzo\inst{\ref{geneve},\ref{oab}}\email{Enrico.Bozzo@unige.ch}
        \and L.~Ji\inst{\ref{zhuhai}}\email{jilong@mail.sysu.edu.cn}
        \and C.~Malacaria\inst{\ref{oar}}\email{christian.malacaria@inaf.it}
        }

   \institute{Institut f\"ur Astronomie und Astrophysik, Kepler Center for Astro and Particle Physics, University of Tuebingen, Sand 1, 72076 T\"ubingen, Germany \label{tue}
   \and Department of Physics, Middle East Technical University, 06800, Ankara, Turkey \label{metu}
   \and Osservatorio Astronomico di Brera, Via E.\ Bianchi 46, I-23807, Merate, Italy \label{oab}
   \and  Department of Astronomy, Universit\'e de Gen\`eve, 16 chemin d'\'Ecogia, 1290 Versoix, Switzerland \label{geneve} 
   \and School of Physics and Astronomy, Sun Yat-Sen University, Zhuhai, 519082, China \label{zhuhai}
   \and INAF Osservatorio Astronomico di Roma, via di Frascati 33, I-00078, Monteporzio Catone, Roma, Italy\label{oar}
   }

   \date{Received Month Day, 2026}

  \abstract
      {We present an analysis of a long \nustar\ observation of the intermediate
        supergiant fast X-ray transient (SFXT) IGR\,J00370+6122, obtained around the peak of its orbital modulation. The observation has a net exposure $\sim$93\,ks.
        During the observation, 
        the source displayed a bright flaring event that was resolved into numerous short flares with durations of $10-10^3$\,s. The $3-70$\,keV luminosity spanned a dynamic range of $\sim$200, from 
        $\sim10^{34}$ to $\sim1.8\times10^{36}$\,erg\,s$^{-1}$; the maximum was reached for only $\sim40$\,s in total. 
        The spectrum is well described by an absorbed double-Comptonisation model, with parameters consistent with those 
        found in other X-ray pulsars accreting at similarly low luminosities. This is the first detection of a double-hump continuum in \src, which extends this spectral state to the class of intermediate SFXTs.
We searched for the previously reported $\sim346$\,s and $\sim674$\,s periodicities and find no significant signal at either period,
which makes their interpretation as the neutron star spin period less likely.
Using $\sim21$\,yr of \emph{Swift}/BAT monitoring, we refine the orbital period of the system to $P_{\rm orb}=15.6654\pm0.0008$\,d. 
Finally, we discuss the observed flaring activity in the framework of quasi-spherical settling accretion, which is expected at the observed luminosities.}

   \keywords{accretion -- stars: neutron -- X-rays:individual IGR J00370+6122}

   \maketitle
\nolinenumbers

\section{Introduction} \label{sect:intro}

High-mass X-ray binaries (HMXBs) consist of an OB-type donor star and a compact object, either a neutron star (NS) or, more rarely, a black hole,
that accretes matter from its companion and produces X-ray emission \citep[for a review, see e.g.][]{Kretschmar19}.
Within this class, supergiant X-ray binaries (SGXBs) host a compact object that orbits an OB supergiant donor and accretes from its powerful, radiatively driven stellar wind.
Classical persistent SGXBs (e.g. Vela\,X$-$1 and 4U\,1700$-$377) are characterised by continuous X-ray emission with moderate variability,
typically about one order of magnitude, which is generally attributed to intrinsic clumpiness and inhomogeneities in the wind from the donor star \citep[e.g.][]{Fuerst10}.

The discovery rate of SGXBs increased substantially following the 2002 launch of the International Gamma-Ray Astrophysics Laboratory (\integral; \citealt{Winkler03}). Its hard X-ray monitoring capabilities revealed a previously unrecognised subclass of highly variable sources, subsequently classified as supergiant fast X-ray transients \citep[SFXTs;][]{Negueruela06,Sguera05,Smith04}. 
Unlike classical SGXBs, SFXTs spend most of their time in a faint, quiescent state ($L_{\rm X}\sim10^{32}-10^{33}$\,erg\,s$^{-1}$),
interrupted by rare, bright outbursts lasting a few $10^{4}$\,s, during which the luminosity typically reaches $L_{\rm X}\sim10^{36}-10^{37}$\,erg\,s$^{-1}$ \citep[e.g.][]{Romano15,Sidoli17}; a few exceptional cases reach up to $\sim10^{38}$\,erg\,s$^{-1}$ (e.g. IGR~J17544$-$2619; \citealt{Romano15b}).
These outbursts are composed of multiple short flares with durations of $\sim1000$\,s.
On short timescales ($<10^{4}$\,s), SFXTs display dynamic ranges in luminosity of $\sim10^{3}-10^{5}$, much larger than those of classical SGXBs,
which makes them one of the most extreme classes of variable HMXBs known.
To date, several dozen SFXTs and SFXT candidates have been identified, although orbital and spin periods
have been measured for only a subset of these sources \citep[e.g.][]{Romano15,Sidoli17}.

The physical mechanism responsible for the extreme variability of SFXTs remains debated, and more than one mechanism likely contributes across the class.
Proposed explanations include (i) clumpiness of the supergiant wind, whereby accretion of individual dense clumps produces sudden
X-ray flares \citep[e.g.][]{intZand05};
(ii) centrifugal and magnetic gating mechanisms, in which the NS spin and its strong magnetic field can temporarily inhibit accretion \citep[e.g.][]{Grebenev07,Bozzo08};
and (iii) quasi-spherical settling accretion, in which a hot, quasi-static shell forms above the magnetosphere and regulates the plasma entry rate through Rayleigh-Taylor instabilities.
In this scenario, moderate flares arise when local fluctuations in the wind density or velocity shift the cooling regime near the magnetosphere, 
while brighter flares occur when the NS captures a magnetised wind structure. This triggers magnetic reconnection, which induces the unstable collapse of the shell onto the NS surface \citep{Shakura14}.

Between the two extremes of classical SGXBs and highly variable SFXTs lies a small group of systems commonly referred to as `intermediate SFXTs'.
These sources show qualitatively similar flaring activity to that of SFXTs, but with a considerably smaller dynamic range, typically of order $\sim10^{2}$ \citep{Walter07}.

\src\ is one such intermediate SFXT, which was discovered in December 2003 with \integral\ in the 20$-$60\,keV \citep{denHartog04} energy range.
The All-Sky Monitor (ASM) on board the Rossi X-ray Timing Explorer (RXTE) ($1.5-12$\,keV) revealed a periodic modulation of the X-ray flux with a period of $15.665 \pm 0.006$\,d,
which was interpreted as the orbital period of the binary system \citep{denHartog04}. 
\citet{Uchida21} later refined the orbital period to $15.6649 \pm 0.0014$\,d (90\% c.l.) using 15\,yr of \emph{Swift}/Burst Alert Telescope (BAT) monitoring data.
Radial velocity measurements have established the orbital parameters of the system.
In particular, \citet[][]{Grunhut14} and \citet[][]{Gonzalez14} derived eccentricities of
$e= 0.48{+0.02\atop-0.03}$ and $e= 0.56 \pm 0.07$, respectively.
Recent measurements based on \emph{Gaia} data place the system at a distance of approximately $3.4\pm0.2$\,kpc \citep{BailerJones21}.

The optical counterpart, BD\,+60$^\circ$\,73, has been the subject of several spectroscopic investigations.
\citet{Gonzalez14} classified it as a BN0.7Ib based on high-resolution spectra and model atmosphere analysis.
\citet{Hainich20} confirmed the main parameters of the optical donor through a detailed UV and optical spectral analysis using the Potsdam Wolf-Rayet atmosphere models.
\citet{Gonzalez14} found that BD\,+60$^\circ$\,73 showed  nitrogen enrichment, indicating an advanced evolutionary stage. They also classified it as an unusual low-luminosity supergiant.

The nature of the compact object in \src\ has been investigated through timing analyses.
\citet{intZand07} detected a possible $346 \pm 6$\,s periodicity during a bright X-ray flare
with the RXTE/Proportional Counter Array (PCA), which they interpreted as the spin period of
an accretion-powered pulsar. However, because the flare was short, the signal repeated only seven times, making this detection uncertain. Subsequently, \citet{Uchida21} analysed \xmm\ data, finding evidence for a $674 \pm 15$\,s periodicity; they propose that the $\sim 346$\,s signal is the second harmonic of this fundamental period.
Nevertheless, this revised periodicity also remains unconfirmed because the signal is transient and appears predominantly during bright flaring states, with only eleven cycles observed.

In the X-ray band, \src\ shows a luminosity ranging from $\lesssim 10^{32}$\,erg\,s$^{-1}$ to $\sim 10^{36}$\,erg\,s$^{-1}$ \citep{Uchida21}.
The X-ray spectral properties have been studied across a broad energy range. \citet{intZand07} modelled the $3-60$\,keV flare spectrum
with an absorbed power law ($\Gamma \approx 2$, $N_{\rm H}\approx 9\times 10^{22}$\,cm$^{-2}$).
\citet{Uchida21} report a harder-when-brighter trend in the $1-10$\,keV band.
The combined RXTE/PCA and INTEGRAL Soft Gamma-Ray Imager (ISGRI) $10-80$\,keV spectrum is well described by a single power law ($\Gamma \approx 2$) with no clear high-energy cutoff. 
The emission is strongly modulated by the orbital phase: the peak occurs after periastron passage, with a delay of about 0.05 in orbital phase,
and the luminosity drops suddenly from $\sim 4\times 10^{33}$\,erg\,s$^{-1}$ to $\sim 10^{32}$\,erg\,s$^{-1}$ at an orbital phase of 0.3 \citep{Uchida21}.

\citet{Gonzalez14} suggested that \src\ is an intermediate SFXT because its X-ray flares resemble those of SFXTs,
but its flux varies over a considerably smaller dynamic range  ($\sim 10^2$) than that of canonical SFXTs ($\gtrsim 10^3$; e.g. \citealt{Romano15,Sidoli17}).
Intermediate SFXTs are generally interpreted as bridging the gap between highly variable, flare-dominated SFXTs and persistently bright, wind-fed classical SGXBs. Fewer than ten intermediate SFXTs are known; two other examples are IGR\,J18483$-$0311 and IGR\,J17354-3255 \citep{Rahoui08, Sguera11}.

In this paper, we present an analysis of a long \nustar\ observation of \src\ that we carried out in 2026 around the expected X-ray peak of the orbital modulation.
The observation lasted about 1.8\,d, with a net exposure time of about 93\,ks. We report all uncertainties at the 1$\sigma$ confidence level unless otherwise stated.

\section{Data analysis} \label{sect:data analysis}

\subsection{NuSTAR}

The Nuclear Spectroscopic Telescope Array (\nustar) satellite is equipped with two identical co-aligned telescopes, each with a focal plane module (FPMA and FPMB).
Both modules operate in the 3$-$79\,keV energy band \citep{Harrison13}. The \nustar\ satellite observed \src\ on 1 March 2026 for a net exposure time of approximately 93\,ks (ObsID 31101014002).
We reduced the data using NuSTARDAS v2.1.5, which is part of HEASOFT v6.35.1, and the calibration files distributed with CALDB v20260309. 

We extracted source events from a circular region centred on the source.
We optimised the extraction radius to maximise the signal-to-noise ratio in the hard part of the X-ray spectrum,
where the source flux is lower and the background contribution is more significant. To achieve this, we used the \texttt{nustar-gen-utils} package (Madsen et al. 2023)\footnote{\url{https://github.com/NuSTAR/nustar-gen-utils}}.
The analysis yielded an extraction radius of $50\arcsec$ centred on the source coordinates.
We extracted background events from a circular region with a radius of $80\arcsec$ centred at 
RA\,$=$\,$00^{\rm h}36^{\rm m}46\fs57$ and 
Dec\,$=$\,$+61\degr24\arcmin45\farcs6$ (J2000),
located on the same detector in a source-free area. We also tested different extraction radii for the source and background
and verified that the spectral results are consistent. We corrected the event times from the satellite frame to the Solar System barycentre
using the \texttt{barycorr} tool and the \texttt{JPLEPH.440} ephemeris.

The \nustar\ data are not continuous. Regular gaps, typically lasting about 21 minutes, are caused by Earth occultations of the source during the satellite's low-Earth orbit, whereas shorter, irregular gaps are introduced by standard GTI screening, which removes passages through the South Atlantic Anomaly and intervals affected by other instrumental constraints.

\begin{figure}
     \centering
     \includegraphics[width=\columnwidth]{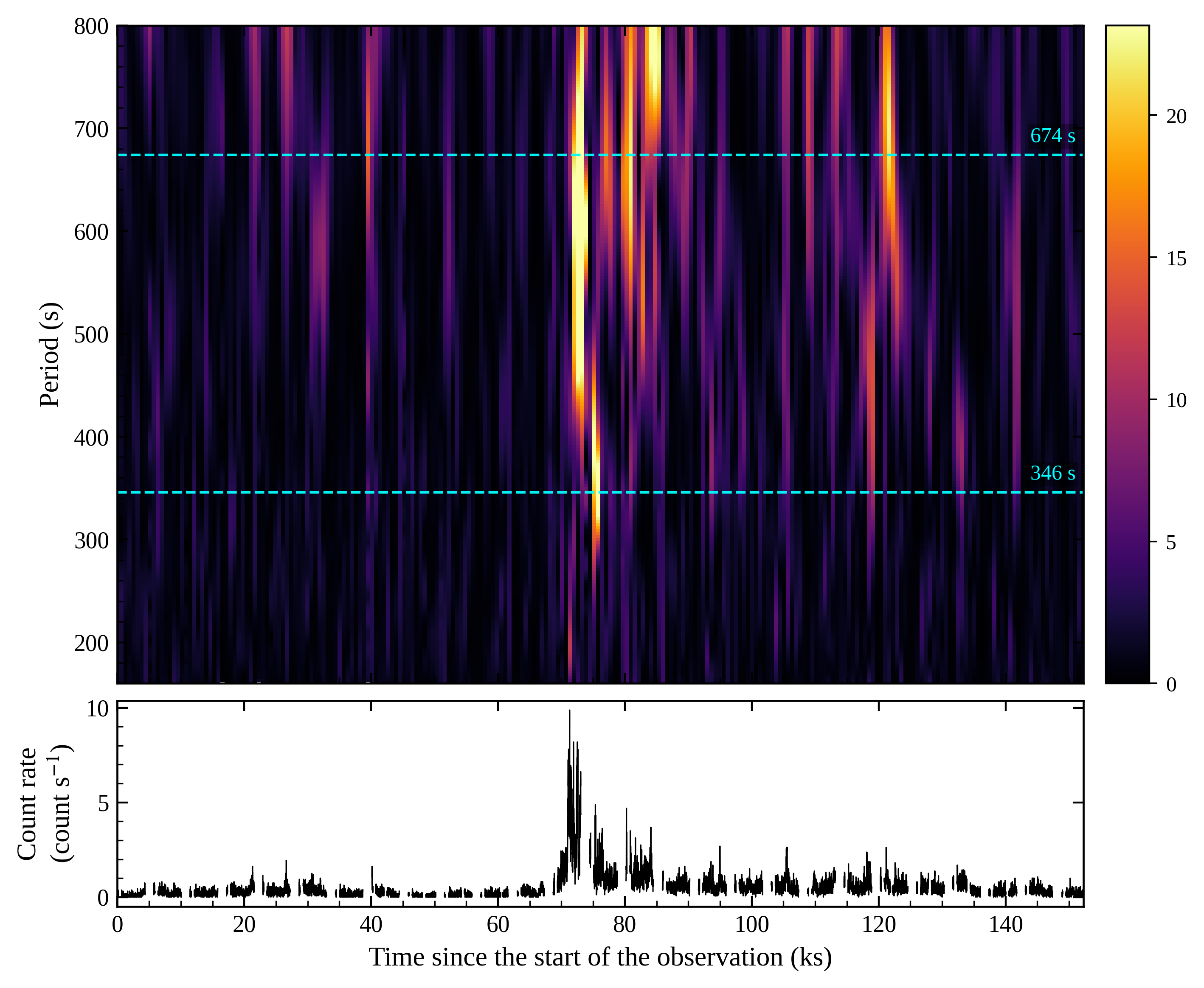}
     \caption{Weighted wavelet transform map of the combined FPMA and FPMB analysis (top panel) and the FPMA and FMPB light curves in the 3$-$20\,keV energy band (bottom panel).}
     \label{fig:wwtmap}
\end{figure}

We searched for the spin period in the 3$-$70\,keV events of modules A and B (combined and separated) using a $Z^2$ test \citep[e.g. ][]{Buccheri83} with up to three harmonics.
We restricted the search to periods between 1\,s and $4\times 10^4$\,s.
We detected no statistically significant pulsations, including in the previously reported periods of $\sim 674$\,s and $\sim 346$\,s.
In addition, given previous reports that the tentative spin-periods were transient and lasted only several cycles \citep{intZand07,Uchida21}, we employed the weighted wavelet transform \citep[WWT;][]{Foster96} to track transient modulations that may not remain stationary throughout the entire observation. For the WWT, we used barycentre-corrected FPMA and FPMB data in the 3$-$20\,keV range (PI channels 35$-$460) with 8\,s binning. We obtained the transforms for both detectors independently and combined them using their geometric mean; we generated the WWT maps in the 160$-$800\,s range (Fig. \ref{fig:wwtmap}). 
The resulting WWT map indicates that the strongest variabilities are concentrated mainly during the bright part of the observation ($\sim 70-85$ ks after the start). Transiently enhanced WWT powers are also present around 674\,s and 346\,s; however, they appear to form part of the formation of several-hundred-second-long, evolving structures rather than a horizontal ridge, as would be expected from coherent pulsations. Thus, the WWT map does not reveal a stable coherent periodicity at $\sim$346 and $\sim$674\,s. These broad structures can more naturally be attributed to sporadic variabilities due to flaring episodes.

To better highlight the variability properties of the source, we re-binned the background subtracted light curve using the optimal segmentation technique based on the Bayesian-block representation described by \citet{Scargle13}\footnote{\url{https://docs.astropy.org/en/stable/api/astropy.stats.bayesian_blocks.html}}.
We applied this algorithm to the event lists to partition the data into an optimal number of blocks, where each block represents a statistically significant change in the count rate. We adopted a false-positive rate probability of $p_0=0.01$ for the Bayesian-blocks analysis to remain sensitive to the rapid rate fluctuations exhibited by the flares.
Because the algorithm accounts for the exposure of each good time interval, it does not merge blocks separated by a gap, even when their count rates are consistent within the uncertainties. For this reason, the light curve shown in Fig.~\ref{fig. lcr} contains consecutive, but not contiguous, blocks with comparable rates.

We extracted the FPMA and FPMB spectra from the same regions used for the timing analysis. We rebinned the \nustar\ spectra to a minimum number of 30 counts per bin so that $\chi^2$ could be used as the fit statistic. We performed the spectral analysis  using XSPEC v12.15.0 \citep{Arnaud96}. 
We included a free multiplicative constant in the spectral fitting to account for calibration uncertainties between the FPMA and FPMB modules.
We modelled the column density ($N_{\rm H}$) with the \texttt{tbabs} model in XSPEC
and we assumed the interstellar-medium abundances (\texttt{wilm} in XSPEC; \citealt{Wilms00}) and \texttt{vern} cross sections \citep{Verner96}.

\subsection{Swift/BAT}

The orbital period of \src\ is known to be $\sim 15.66$\,d, although slightly different values have been reported in the literature.
The most recent determination, based on approximately 15\,yr of \textit{Swift}/BAT observations, is $15.6649 \pm 0.0014$\,d  at a 90\% confidence level \citep{Uchida21}. We refined the orbital period using the full \textit{Swift}/BAT monitoring available to date, covering approximately 21\,yr.
We retrieved the $15-50$\,keV orbital light curve of \src\ from the \textit{Swift}/BAT Transient Monitor\footnote{\url{https://swift.gsfc.nasa.gov/results/transients/}}
\citep{Krimm13}. The light curve is averaged over the \emph{Swift} orbital period of $\sim90$ minutes and covers the interval between MJD 53416 and 61068. We excluded all measurements with quality flags 1, 2, and 3. We searched for periodic signals using the Lomb-Scargle periodogram \citep{Lomb76, Scargle82} implemented in the \texttt{astropy} Python package\footnote{\url{https://docs.astropy.org/en/stable/timeseries/lombscargle.html}}. We adopted the fast $\chi^2$ algorithm of \citet{Palmer09} with three Fourier terms. We searched over the $2-1000$\,d period range, oversampling the independent frequency spacing by a factor of 20 to improve the determination of the periodogram peak.
We estimated the uncertainty on the orbital period through 1000 Monte Carlo simulations.
We generated synthetic light curves by randomly sampling each data point according to its measured statistical uncertainty,
assuming Gaussian errors. We adopted the standard deviation of the recovered orbital periods as the uncertainty.

\section{Results} \label{sect:results}

   \begin{figure}
     \centering
     %
     \includegraphics[width=\columnwidth]{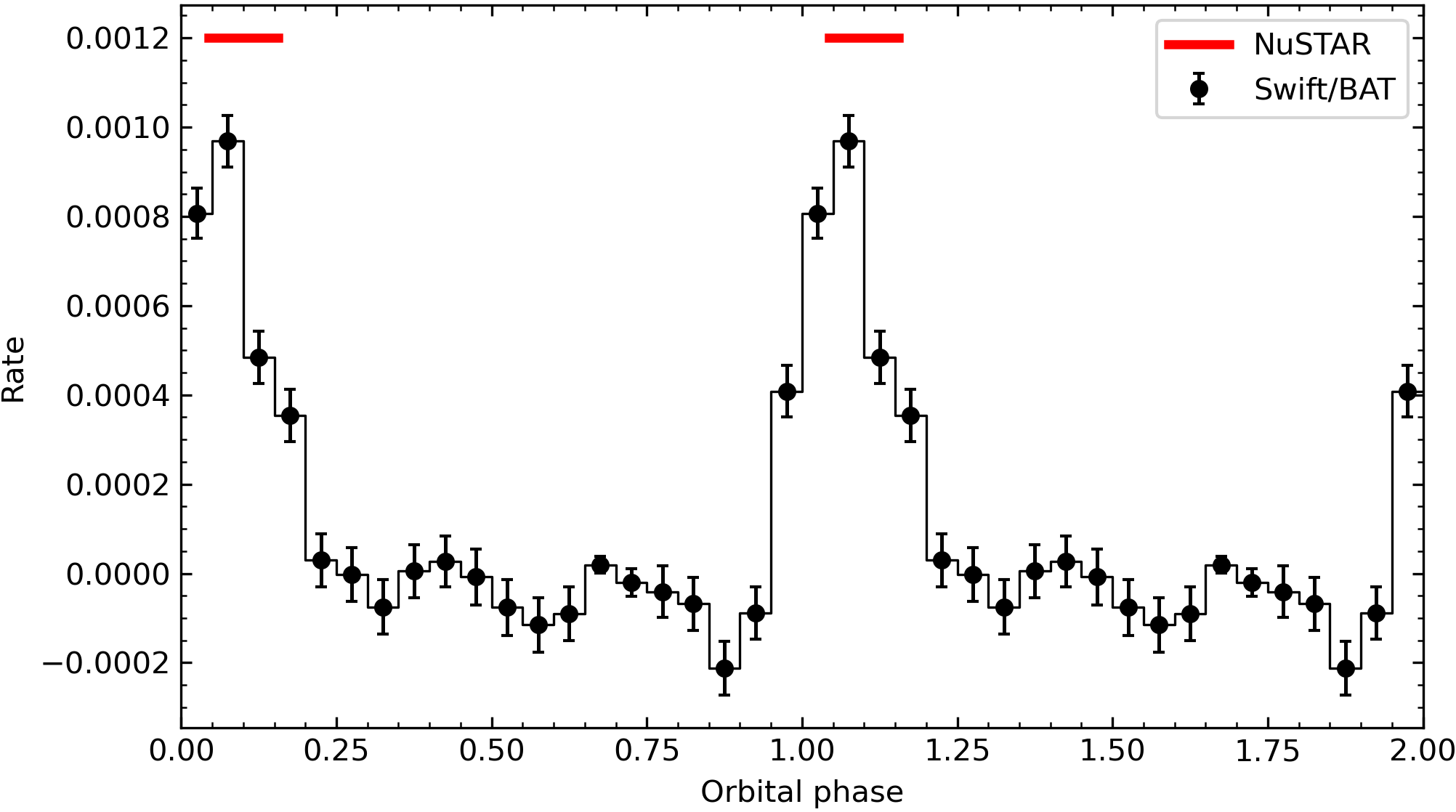}
   \caption{Folded \emph{Swift}/BAT light curve of \src\ using the orbital period $P_{\rm orb} =  15.6654 \pm 0.0008$\,d and $T_0 =55084.02$\,MJD (phase 0 and periastron passage according to \citealt{Gonzalez14}).
The horizontal red line indicates the orbital-phase interval covered by the \nustar\ observation.}
   \label{fig:folded_lightcurve}
   \end{figure}

\subsection{Orbital period}

We determined the orbital period of \src\ by applying the Lomb-Scargle periodogram to the full \textit{Swift}/BAT light curve.
The resulting value is $P_{\rm orb} = 15.6654 \pm 0.0008$\,d, where the uncertainty is the $1\sigma$ standard deviation.
This is currently the most accurate orbital-period measurement for this source and improves on the previous measurement by \citet{Uchida21}.
We folded the \emph{Swift}/BAT light curve using this orbital period
and adopted a reference epoch $T_0 =55084.02$\,MJD (phase 0), which corresponds to the periastron passage calculated  by \citet{Gonzalez14}. 
Figure~\ref{fig:folded_lightcurve} shows the folded light curve. The horizontal red lines in the same figure mark
the orbital-phase interval covered by the \nustar\ observation.

   \begin{figure*}
     \centering
     \includegraphics[width=14cm]{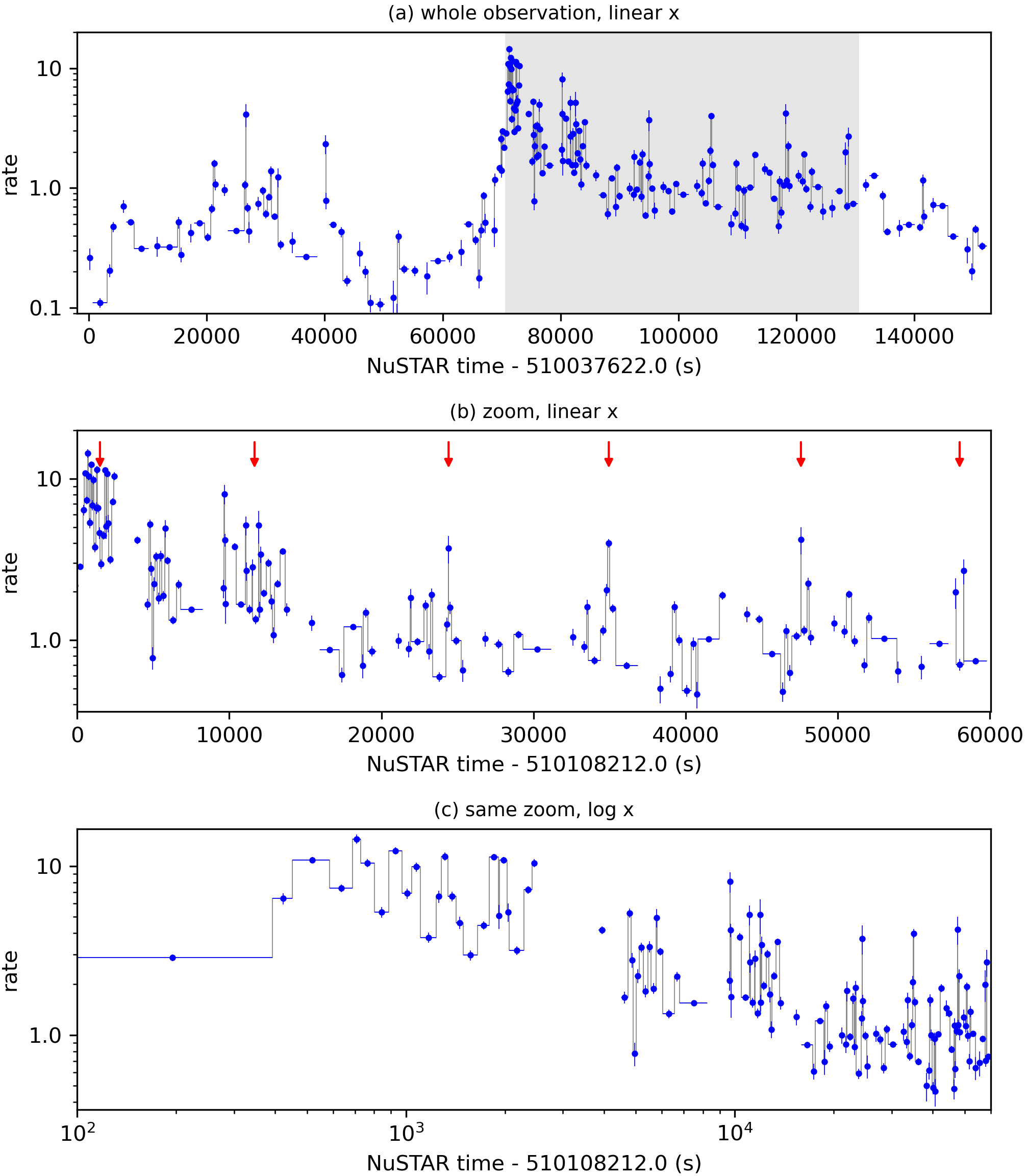}
     \caption{Light curve of \src\ observed with \nustar\ in the $3-70$\,keV energy band, obtained using  Bayesian block segmentation. \emph{Top panel}: Full light curve.
     \emph{Middle panel}: Zoomed-in view of the shaded region in the top panel, with a linear time axis.
    \emph{Bottom panel}: Same zoomed-in view, with a logarithmic time axis.
    In all panels the grey line represents a step function through the Bayesian blocks;
    the line is interrupted at data gaps caused by Earth occultations
    and other GTI filtering criteria. In the middle and bottom panels, time is
    measured from the beginning of the shaded region. The
    linear axis of the middle panel preserves the flare durations
    throughout the outburst, while the logarithmic axis in the bottom panel
    expands the first few kiloseconds, making the short flares around the outburst peak more visible. The six red arrows in the middle panel mark the peaks whose spacing is discussed in Sect.\,\ref{sect:discussion}.}
   \label{fig. lcr}
   \end{figure*}

\subsection{X-ray variability}

The top panel of Fig. \ref{fig. lcr} shows the $3-79$\,keV light curve, obtained with the optimal segmentation technique based on the Bayesian-block representation (see Sect. \ref{sect:data analysis})
 This light curve displays a strong flaring activity.
During the first approximately 40\,ks of the observation, \src\ shows mild flaring activity,
with a maximum flux ratio of $F_{\rm max}/F_{\rm min}$ of approximately $40$. After this period, the flux decreases significantly,
dropping by a factor of approximately ten for roughly 20\,ks.
This decline is followed by a brighter phase of flaring activity,
as shown in the bottom panel of Fig.\,\ref{fig. lcr}. Hereafter, we refer to this episode as the outburst.

   \begin{figure}
     \centering
          \includegraphics[width=9cm]{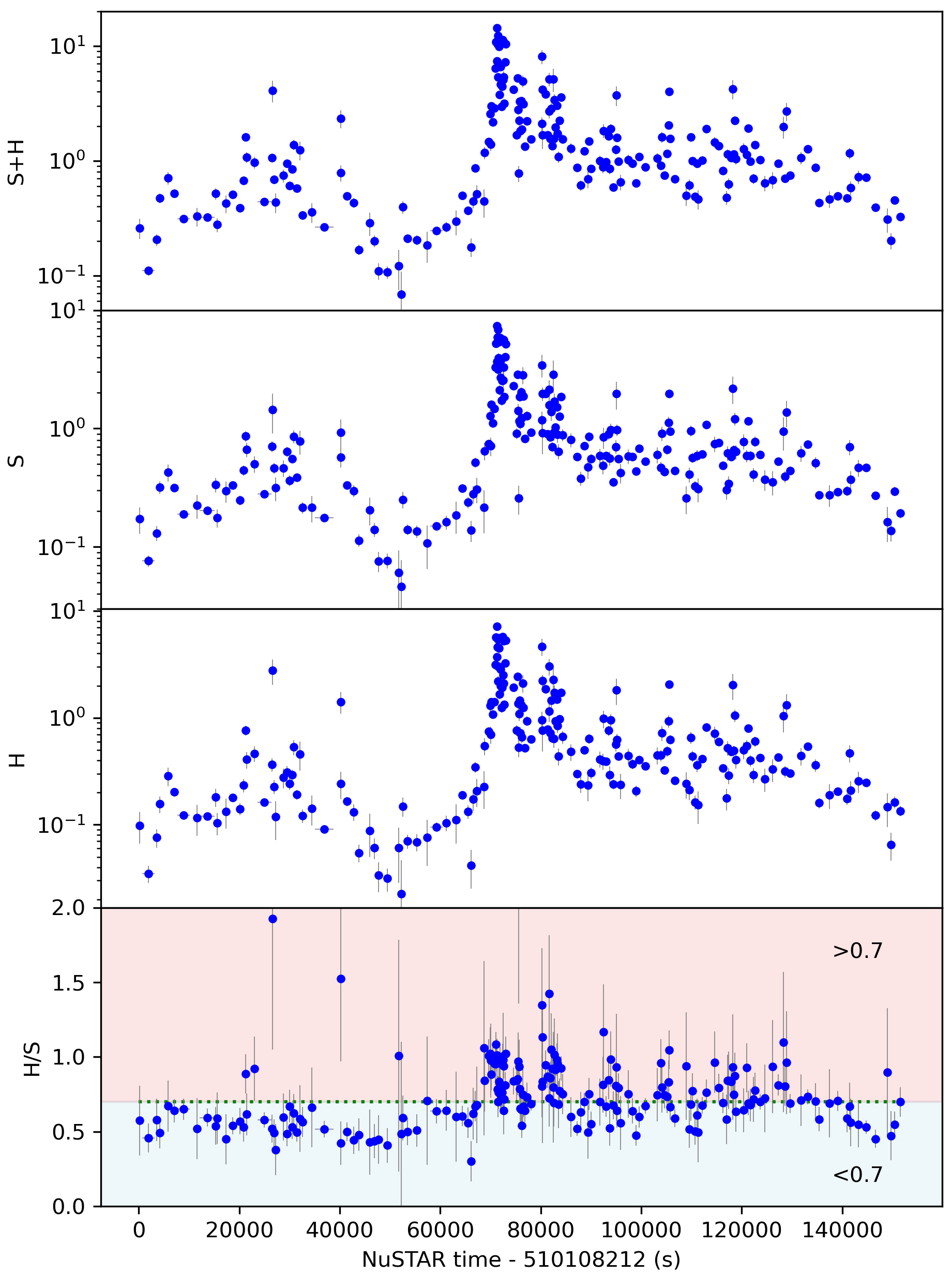}\\
          \includegraphics[width=9cm]{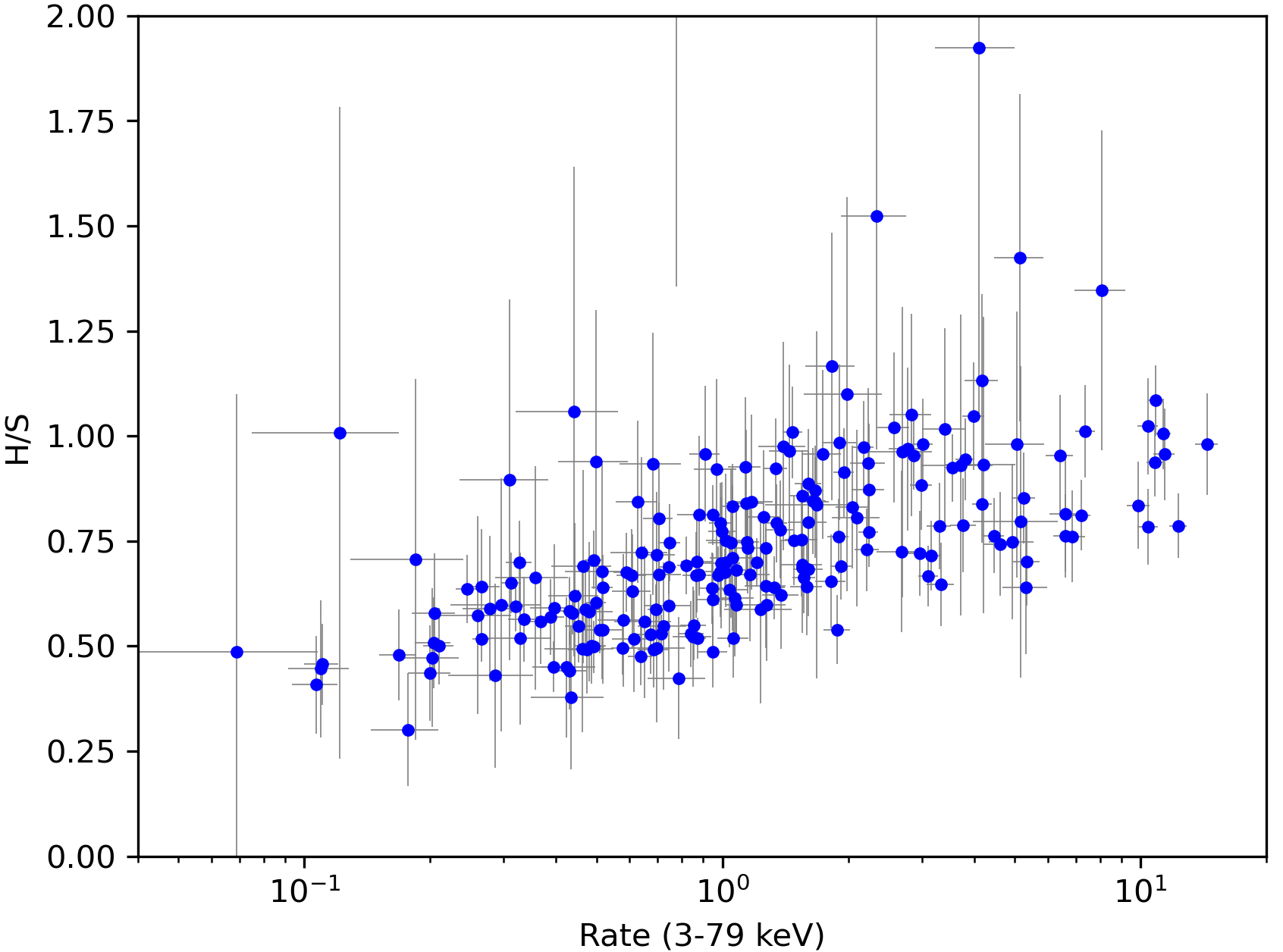}
   \caption{\emph{Main panel}: From top to bottom, \nustar\ light curves in the total (H+S: 3$-$79\,keV), soft (S: 3$-$7\,keV), and hard (H: 7$-$79\,keV) bands, and the hardness ratio, all shown as a function of time. \emph{Bottom panel}: Hardness ratio as a function of the total 3$-$79\,keV count rate.}
   \label{fig:hr}
   \end{figure}

\subsection{Spectral variability} 

To investigate the spectral variability of \src\ during the \nustar\ observation 
we constructed hardness-ratio (HR) light curves by extracting source counts in the soft ($3-7$\,keV) and hard ($7-79$\,keV) energy bands.
The top panel of Fig.~\ref{fig:hr} shows the total ($3-79$\,keV), soft, and hard light curves together with the evolution of the HR as a function of time,
while the bottom panel displays the HR as a function of the total X-ray flux.
The HR analysis reveals mild spectral variability throughout the observation, with the source generally becoming harder at higher flux levels.

To investigate this behaviour further, we performed HR-resolved spectroscopy by extracting two separate spectra based on the hardness ratio:
one comprising data with HR\,$\leq 0.7$ (hereafter, the soft spectrum) and the other with HR\,$> 0.7$ (the hard spectrum).
We chose the threshold value of HR\,$=0.7$ to divide the data into two spectra with comparable statistics.

For the continuum, we tested several models commonly used to describe the spectra of accreting compact objects 
in HMXBs,
including a cutoff power law and a power law with a high-energy exponential cutoff.
We find that the best fit, based on the $\chi^2$ statistic and the residuals, is obtained with a double-Comptonisation model
(\texttt{comptt} + \texttt{comptt} in XSPEC; \citealt{Titarchuk94}). This model has been  used successfully to describe the broadband spectra
of several X-ray pulsars observed at low luminosities ($L_{\rm X}\approx 10^{33}-10^{35}$\,erg\,s$^{-1}$),
including X\,Per \citep{Doroshenko12}, GX\,304$-$1 \citep{Tsygankov19a}, A\,0535+262 \citep{Tsygankov19b}, 
2RXP\,J130159.6$-$635806 \citep{Salganik25},  MAXI\,J0655$-$013 \citep{Malacaria26},
RX\,J0440.9+4431 \citep{Salganik23}, and 1A\,1118$-$61 \citep{Salganik26}. 
Interestingly, in the latter two sources, the double hump is seen up to X-ray luminosities of $\sim10^{36}$\,erg\,s$^{-1}$. 
In all the low-luminosity accreting pulsars mentioned above, the double-\texttt{comptt} model
is interpreted as arising from two distinct emission components: a soft component originating from thermal Comptonisation in the accretion-heated polar cap region,
and a hard component associated with either cyclotron emission reprocessed by magnetic Compton scattering in the overheated upper layers
of the NS atmosphere or thermal Comptonisation in a hotter plasma (\citealt{Tsygankov19a, Tsygankov19b}, and references therein).

   \begin{figure*}
     \centering
     \includegraphics[width=\columnwidth]{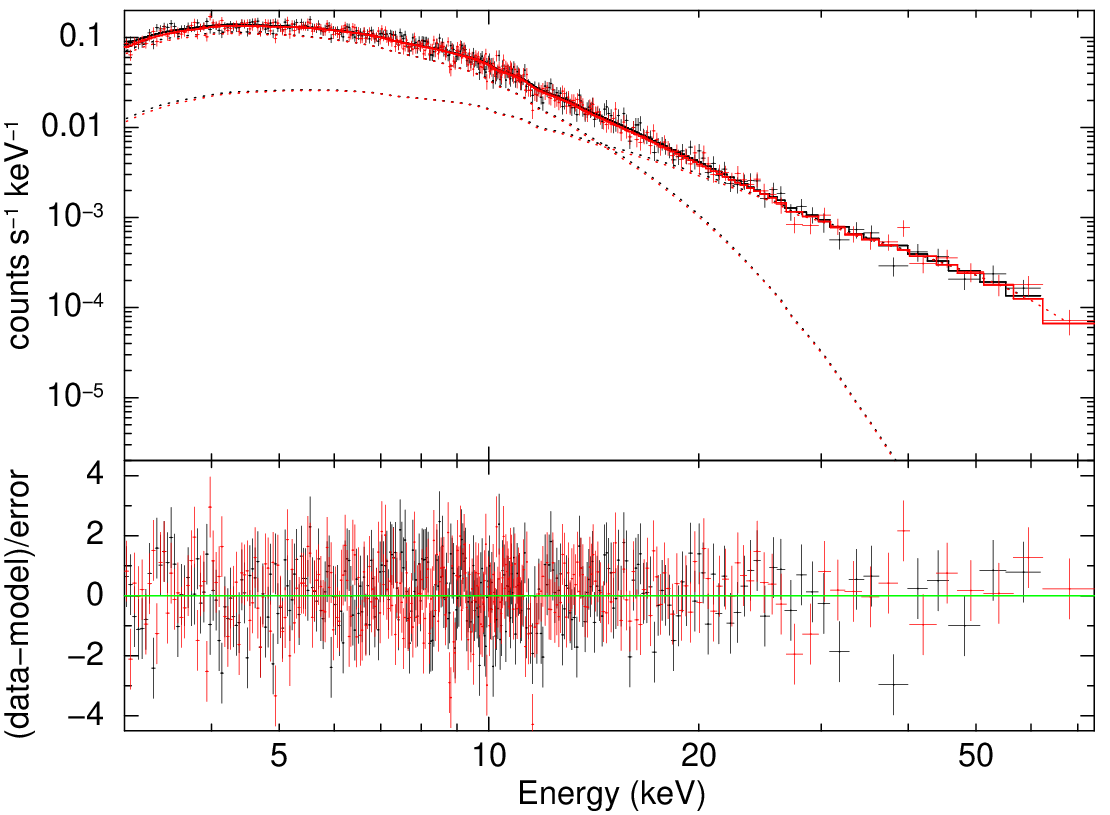}
     \includegraphics[width=\columnwidth]{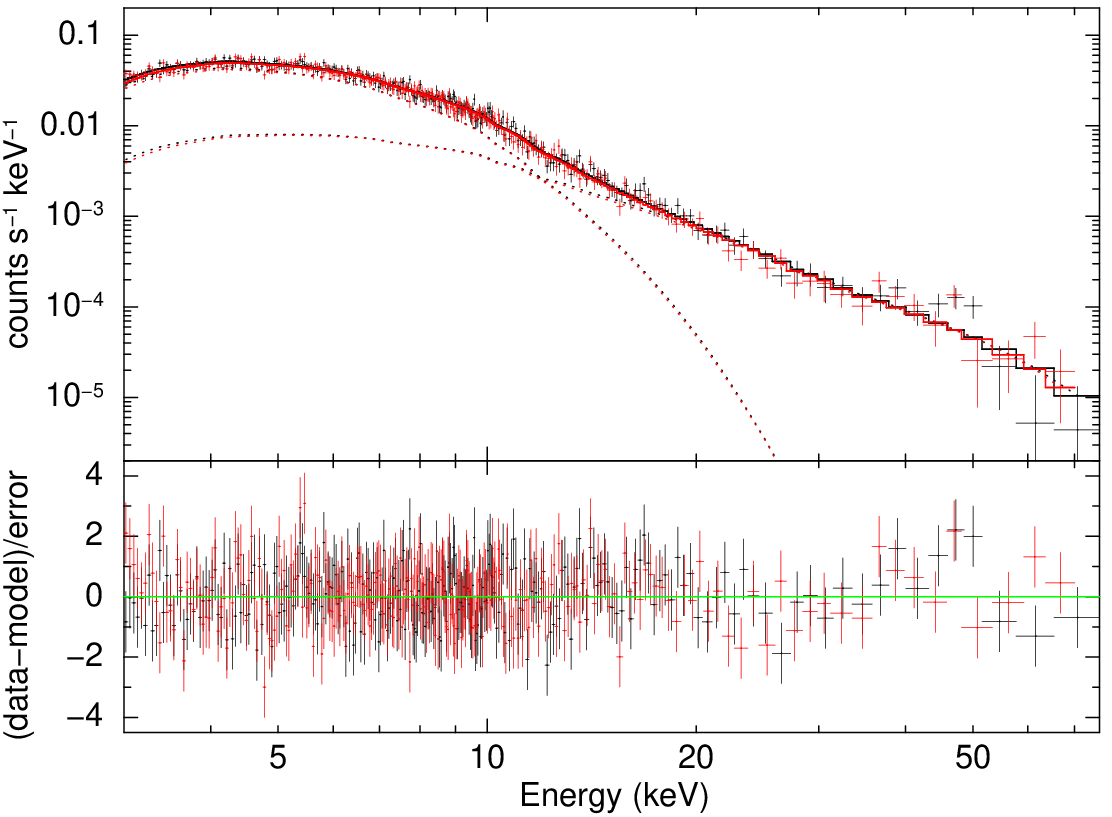}
     \caption{Spectra obtained with \nustar (FPMA in black, FPMB in red) for the hard (HR\,$>0.7$; left panel) and soft (HR\,$<0.7$; right panel) HR selections,
       fitted with the absorbed double-\texttt{comptt} model (solid lines) and its individual components (dashed lines).
       The lower panels show the residuals with respect to the best-fit model.}
   \label{fig. HRs spectra}
   \end{figure*}

\begin{table}
\centering
\caption{Best-fit results for the hard (HR\,$>0.7$) and soft (HR\,$\leq0.7$) spectra
shown in Fig.~\ref{fig. HRs spectra}. \label{tab:hr_params}}
\begin{tabular}{lcc}
\hline
\noalign{\smallskip}  
Parameter & \multicolumn{1}{c}{HR~$>$~0.7} & \multicolumn{1}{c}{HR~$\leq$~0.7} \\
\noalign{\smallskip}
\hline
\noalign{\smallskip}
$N_{\rm H}$  ($10^{22}$\,cm$^{-2}$)  & $4.9{+0.9 \atop -1.0}$ & $4.8{+1.2 \atop -0.9}$ \\
\noalign{\smallskip}
$kT_0$   (keV)    & $0.29{+0.08 \atop -0.08}$ &  $0.31{+0.09 \atop -0.10}$ \\
\noalign{\smallskip}
$kT_{e,1}$ (keV)   & $3.0{+0.3 \atop -0.3}$ & $1.92{+0.20 \atop -0.10}$ \\
\noalign{\smallskip}
$\tau_{\rm 1}$ & $11.8{+1.4 \atop -0.8}$  & $15.4{+1.1 \atop -2.0}$  \\
\noalign{\smallskip}
norm$_1$     & $9.8{+3.5 \atop -1.8}\times 10^{-3}$ & $5.8{+1.7 \atop -1.4}\times 10^{-3}$ \\
\noalign{\smallskip}
$kT_{\rm e,2}$ (keV) & $44{+7 \atop -10}$  &  $30.5{+2.3 \atop -2.4}$  \\
\noalign{\smallskip}
$\tau_{\rm 2}$ & $5.5{+0.8 \atop -0.7}$ &  $5.0{+0.7 \atop -0.6}$ \\
\noalign{\smallskip}
norm$_2$     & $1.3{+0.5 \atop -0.2}\times 10^{-4}$ & $5.2{+1.6 \atop -1.2}\times 10^{-5}$ \\
\noalign{\smallskip}
const$_{AB}$  & $1.045{+0.010 \atop -0.013}$ & $1.037{+0.012 \atop -0.014}$ \\
\noalign{\smallskip}
\hline
\noalign{\smallskip}
$F_{\rm x}$ (erg\,cm$^{-2}$\,s$^{-1}$)   & $(1.022{+0.024\atop-0.010})\times 10^{-10}$ & $(2.69{+0.07\atop-0.06})\times 10^{-11}$ \\
\noalign{\smallskip}
\hline
\noalign{\smallskip}
Total $\chi^2$/d.o.f. & 630.19 / 588 & 456.58 / 510 \\
\noalign{\smallskip}
\hline
\noalign{\smallskip}
\end{tabular}
\ \\
{\small Notes. The quantity $F_{\rm x}$ is the $3-70$\,keV absorbed flux. Both \texttt{comptt} components share the same $kT_0$ value.}
\end{table}

Figure ~\ref{fig. HRs spectra} shows the best-fitting spectra for the soft and hard HR states, and Table~\ref{tab:hr_params} shows the corresponding best-fit parameters.
We modelled the absorption with a single
\texttt{tbabs} component, so the column densities reported in
Table~\ref{tab:hr_params} include both Galactic and local contributions.
Because the Galactic column density in the direction of the source is
$N_{\rm H\,Gal} \approx 4.8 \times 10^{21}$~cm$^{-2}$ \citep{HI4PI16}, approximately an order of
magnitude lower than the measured values, the observed absorption is mainly
intrinsic to the source. Intrinsic column densities of this order are commonly
observed in HMXBs, including SFXTs, where they are usually ascribed to the dense
stellar wind of the supergiant donor in which the compact object is embedded
\citep[e.g. ][]{Kretschmar19}.

An alternative continuum model consisting of an absorbed blackbody plus a power law also provides a reasonable description of the data,
although with worse $\chi^2$ (hard spectrum: $\chi^2/\mathrm{d.o.f.} = 647.44/591$; soft spectrum: $\chi^2/\mathrm{d.o.f.} = 465.43/514$) and less satisfactory residuals.
For this model, we obtained a blackbody temperature of $kT_{\rm bb} \approx 1.4-1.8$\,keV, a blackbody radius of $R_{\rm bb} \sim 130-160$\,m,
a photon index of $\Gamma \approx 2$, and a column density varying from $N_{\rm H} \sim 2.3\times 10^{22}$\,cm$^{-2}$ for the hard spectrum
to essentially zero for the soft spectrum.
Given the better fit quality of the double-\texttt{comptt} model, we adopted it as our preferred description of the continuum.

We used the spectral results presented above to estimate the minimum and maximum
luminosity reached by \src\ during the \nustar\ observation. We converted the lowest and highest count rates of the light curve shown in Fig.~\ref{fig. lcr} into fluxes, using the unabsorbed fluxes of the two best-fitting models reported
in Table~\ref{tab:hr_params}. 
Because the HR correlates with flux (bottom panel of Fig.~\ref{fig:hr}), 
we adopt the HR\,$\leq0.7$ model for the minimum rate and the HR\,$>0.7$ model for the maximum count rate. For an assumed distance of $3.4$\,kpc
\citep{BailerJones21}, we obtain $3-70$\,keV luminosities of $\sim10^{34}$\,erg\,s$^{-1}$ and $\sim1.8\times10^{36}$\,erg\,s$^{-1}$,
corresponding to a dynamic range of approximately 200. When extrapolated
to the $0.1-100$\,keV band, the luminosities derived from the same models are $\sim1.6\times10^{34}$\,erg\,s$^{-1}$ and
$\sim3\times10^{36}$\,erg\,s$^{-1}$.

\section{Discussion}
\label{sect:discussion}

Throughout the \nustar\ observation, \src\ remained below 
$L_{\rm x}\simeq4\times10^{36}$\,erg\,s$^{-1}$, the limit below which 
the shell above the magnetosphere cools inefficiently 
and quasi-spherical settling accretion is expected to operate \citep{Shakura14,Postnov14}.
The luminosity reached during the brightest flares
approaches this limit, but only for a total of about $\sim 40$\,s over the entire exposure, and it remains below the limit even then.
Throughout the observation, \src\ therefore remained in the luminosity regime where a quasi-static shell can form above the magnetosphere and regulate
the accretion onto the NS.
Using the orbital and stellar-wind parameters of \src\ from \citet{Gonzalez14} and \citet{Hainich20},
we therefore checked whether the observed flaring properties are consistent with 
episodes of enhanced plasma entry, triggered by instabilities within the quasi-static shell 
that surrounds the NS magnetosphere.
We also note that the low-state luminosity of \src\ (a few $10^{34}$\,erg\,s$^{-1}$)
is of the order expected for the non-flaring state of an SFXT in the model by \citet{Shakura14},
and that the dynamic range of about two orders of magnitude spanned by
\src\ during the \nustar\ observation is consistent with the flare amplitudes
that can be produced when the magnetospheric plasma entry rate switches from the settling value to free fall.

In the quasi-spherical settling accretion model, a flare develops on the free-fall time-scale from the outer boundary of the shell,
$t_{\rm ff}=R_{\rm acc}^{3/2}/\sqrt{GM_{\rm NS}}$, with
\begin{equation}
  R_{\rm acc}=\frac{2GM_{\rm NS}}{v_{\rm rel}^{2}},\qquad
  v_{\rm rel}^{2}=\left(v_{\rm w}-v_{r}\right)^{2}+v_{\varphi}^{2},
  \label{eq:racc}
\end{equation}
where $v_{\rm w}$ is the radial wind velocity at the orbital separation of the NS, and $v_{r}$ and $v_{\varphi}$ are the radial and
azimuthal components of the orbital velocity \citep[][]{Okuda77}. For $v_{\rm w}$ we adopted the
double-$\beta$ velocity law of \citet{Hainich20}. The orbital terms cannot be
neglected, since the observation covers periastron passage, where $v_{\varphi}$
is comparable to the local wind velocity \citep[see also][]{Hainich20}. Two interpretations are then possible.

1. The outburst as a single shell collapse.
The interpretation of the entire outburst as resulting from shell accretion requires
$t_{\rm ff}\approx 5-7\times10^{4}$\,s and hence a low relative velocity. We used
a terminal wind velocity of $v_{\infty}=900$\,km\,s$^{-1}$, corresponding to the $1\sigma$ lower bound from
\citet{Hainich20}, together with the orbital solution of
\citet{Gonzalez14} to obtain $R_{\rm acc}\approx 3.1\times10^{11}$\,cm and
$t_{\rm ff}\approx 1.3\times10^{4}$\,s, within a factor of approximately five of the observed
duration. However, the stellar
masses and wind parameters of \src\ are poorly constrained, and the adopted wind law
neglects the feedback of the X-ray emission on the wind. Photoionisation
suppresses line driving and would lower $v_{\rm w}$ further below the value assumed here
\citep[e.g. ][]{Ducci10,Bozzo21}. Because $R_{\rm acc}\propto v_{\rm rel}^{-2}$
and $t_{\rm ff}\propto v_{\rm rel}^{-3}$, a modest reduction in $v_{\rm w}$ would suffice to reduce the discrepancy between the observed duration of the outburst and $t_{\rm ff}$.

2. The outburst as a sequence of depletion and replenishment cycles.
Interestingly, the value $t_{\rm ff}\approx1.3\times10^{4}$\,s derived above is comparable to the intervals
between the outburst peak and the five subsequent rebrightenings,
which range between $\sim$10 and $\sim$13\,ks (see the red arrows in the middle panel of Fig.\,\ref{fig. lcr}).
In this scenario, the first collapse of the shell produces the initial peak,
and the subsequent enhancements correspond to further collapses following the 
replenishment of the shell by newly captured wind material.
Such a recurrence is expected in the model by \citet{Shakura14,Postnov14}, in which flares repeat for as long
as the enhanced plasma entry rate is sustained. 
We caution that we may have missed
some sporadic flares during the $\sim1.4$\,ks gaps separating the $\sim4.4$\,ks
visibility windows; the pattern discussed here therefore remains only suggestive.

An important feature of the outburst observed by \nustar\ and reported in this work is that it is composed of numerous
flares with durations of $\sim10-10^3$\,s (Sect.~\ref{sect:results}),
shorter than $t_{\rm ff}(R_{\rm acc})$ under either of the interpretations above.
We speculate that these flares could reflect a non-uniform breakup of the shell, with matter
leaking onto the NS surface from a few preferential regions of the magnetospheric boundary rather than  simultaneously over the $4\pi$ solid angle.
These flares may also explain the $\sim674$\,s and $\sim346$\,s periodicities reported by
\citet{intZand07} and \citet{Uchida21}, which were found only during bright flaring episodes
and over a limited number of cycles. A few consecutive flares with durations and separations of this order can mimic periodicity over a few cycles.


The \nustar\ spectrum of \src\ is well described by an absorbed double-\texttt{comptt} model,
a characteristic feature of X-ray pulsars accreting at low mass accretion rates (typically $L_{\rm x} \approx 10^{33}-10^{35}$\,erg\,s$^{-1}$).
Previous spectral studies of this source were limited to single power-law or
cutoff power-law descriptions \citep{intZand07,Uchida21}. The sensitivity and broad
energy coverage of the present observation reveal the double-hump
continuum in \src\ for the first time. This adds a new member to the small group of sources in which
this spectral state has been resolved and extends this phenomenology to the class of
intermediate SFXTs, where, to our knowledge, it has not previously been reported.
At these low luminosities, X-ray pulsars are in the subcritical accretion regime, in which the emission region is a hotspot (rather than an accretion column) at the NS surface \citep{Mushtukov21}.
This double-hump spectrum is physically interpreted as arising from a strong temperature gradient in the NS atmosphere.
The soft component, which peaks around $5-10$\,keV originates from thermal Comptonisation in the optically thick, deep atmospheric layers of the polar cap, and is
dominated by extraordinary polarised photons that escape due to their reduced scattering cross section. 
The hard component originates from the Comptonisation of the total photon field in the hot upper 
atmosphere. This field consists of thermal photons that escape from the bottom  and cyclotron photons produced
as infalling ions decelerate and collisionally excite electrons, which then undergo multiple scatterings in the hot atmospheric plasma
\citep{Sokolova-Lapa21, Mushtukov21, Tsygankov19a}.
In our analysis, we find that the normalisation of the hard-\texttt{comptt} component tends to increase with luminosity more strongly than that of the soft \texttt{comptt} component,
although the uncertainties on the normalisations do not allow a firm conclusion.
A similar trend has been reported in other accreting pulsars at comparable luminosities, for example, in 1A~1118$-$61 \citep{Salganik26}. 
\citet{Salganik26}  qualitatively interpret this behaviour as the result of resonant reprocessing of atmospheric radiation within the accretion flow above the polar cap: photons emitted near the NS surface with energies close to the local cyclotron energy can scatter resonantly off the infalling Doppler-shifted electrons, exciting them to higher Landau levels whose subsequent decay regenerates photons near the cyclotron energy. This mechanism is also known as photon spawning \citep{Araya99}. Part of this reprocessed radiation is scattered back toward the star and re-emitted, seeding further resonant interactions. 
As the mass accretion rate increases, the resonant optical depth of the infalling column also increases, making this mechanism more efficient and directing more flux into the high-energy hump, while the low-energy component grows more slowly.

\section{Conclusions}

We analysed a long \nustar\ observation of \src, with a net exposure of $\sim93$\,ks. The observation was performed around the peak of its orbital modulation.

The observation was caught \src\ in a bright flaring event, providing the most detailed view
obtained so far of its short-term X-ray variability. The
emission is resolved into numerous short flares with durations of $\sim10-10^3$\,s.

The source exhibited a dynamic range of approximately 200 over the observation, with luminosities ranging from $\sim10^{34}$ to $\sim1.8\times10^{36}$\,erg\,s$^{-1}$.

Throughout the observation, \src\ remained below
$L_{\rm x} \simeq 4\times10^{36}$\,erg\,s$^{-1}$, the luminosity at which
quasi-spherical settling accretion is expected to operate. 
We therefore examined  the flaring activity within this framework. Depending on the adopted wind velocity, the free-fall time from the outer boundary of the shell is
comparable either to the duration of the entire outburst, which would then
correspond to a single collapse of the shell, or to the duration of its initial
bright phase, in which case the subsequent enhancements would be produced by
successive cycles of depletion and replenishment of the shell. The sampling of
the present observation does not allow us to distinguish between these two
possibilities. In both cases, the flares are much shorter than the free-fall
time. We suggest that they reflect a non-uniform break-up of the shell,
with matter entering the magnetosphere from a few preferential regions rather
than simultaneously over the whole magnetospheric boundary.

We searched for the $\sim346$\,s and $\sim674$\,s periodicities reported in the
literature, using both a $Z^2$ search and a weighted wavelet transform, but
found no significant signal at either period. This makes their
interpretation as the spin period of the NS less likely and is consistent with
the transient nature already noted by \citet{intZand07} and \citet{Uchida21}.

The \nustar\ spectrum is well described by an absorbed double-Comptonisation
model. We report the first detection of a double-hump continuum in \src. This spectral state has previously been observed only in other types of X-ray
pulsars accreting at low rates, and our results extend it to the class of intermediate SFXTs. The measured
column density is about an order of magnitude higher than the Galactic value in
the direction of the source, indicating that the observed absorption is mostly intrinsic.

Finally, using $\sim21$\,yr of \swift/BAT monitoring, we refine the orbital
period of the system to $P_{\rm orb} = 15.6654 \pm 0.0008$\,d, which is
currently the most accurate measurement available for this system.

We show that the properties of the flaring activity reported here are compatible with
quasi-spherical settling accretion. 
To discriminate this scenario from magnetic or centrifugal gating mechanisms, the corotation radius and the size of the magnetosphere, 
which can be inferred from  measurements of the NS spin period and magnetic field, are still required.
Because the orbital parameters of \src\ are already well determined, once these two quantities 
are measured, the system would become one of the few SFXTs in which most of the parameters describing these models are known,  and hence one of the most suitable targets for testing them.

\begin{acknowledgements}
We thank the anonymous referee for helpful comments that improved this paper.
LD acknowledges funding from the Deutsche Forschungsgemeinschaft (DFG, German Research Foundation) - Projektnummer 549824807.
PR and EB acknowledge financial support from the Bando di Finanziamento della Ricerca Fondamentale INAF 2023 Mini Grant “Investigating SFXTs and other wind-fed HMXBs”. 
CM was supported by Fondazione Cariplo/Cassa Depositi e Prestiti (Grant 2023-2560, PI: Papitto).
This research has made use of the NuSTAR Data Analysis Software (NuSTARDAS) jointly developed by the ASI Space Science Data Center (SSDC, Italy) and the California Institute of Technology (Caltech, USA).
This research made use of \textsc{Astropy}, a community-developed core Python package for Astronomy \citep{astropy:2013, astropy:2018, astropy:2022}, and of \textsc{Stingray} \citep{Huppenkothen19, Bachetti24, matteo_bachetti_2024_11383212}, a community-developed spectral-timing software package for astrophysics.
\end{acknowledgements}  

\bibliographystyle{aa} 
\bibliography{aa62434-26}

\end{document}